\documentstyle[12pt]{article}
\begin{document}

\title{Exact Solution Versus Gaussian Approximation for a
 Non-Ideal Bose Gas in One-Dimension}

\vskip 1.0cm

\author{P.R.I. Tommasini \\
Institute for Theoretical Atomic and Molecular Physics \\ 
Harvard-Smithsonian Center for Astrophysics \\ 
Cambridge, MA 02138 \\ \\
P.L. Natti \\
Departamento de Matem\'atica, Universidade 
Estadual de Londrina \\
C. P. 6001, 86051-990, Londrina, PR, Brazil\\ \\
E.R. Takano Natti \\
Universidade do Norte do Paran\'a \\
Av. Paris, 675 , 86041-140, Londrina, PR, Brazil \\ \\
A.F.R. de Toledo Piza and C.-Y. Lin\\
Instituto de F\'{\i}sica, Universidade de S\~ao Paulo, \\
05389-970, S\~ao Paulo, S.P. , Brazil}

\maketitle

\vskip 0.3cm
\centerline{\large{\it submitted to Physica {\bf A}}}
\newpage

\begin{abstract}
\noindent 
We investigate ground-state and excitation spectrum of a
system of non-relativistic bosons in one-dimension 
interacting through repulsive, two-body contact interactions 
in a self-consistent 
Gaussian mean-field approximation. The method 
consists in writing the 
variationally determined density operator 
as the most general Gaussian 
functional of the quantized field operators. 
There are mainly two advantages in working 
with one-dimension. First, the 
existence of an exact solution for the 
ground-state and excitation energies. 
Second, neither in the perturbative results nor in the 
Gaussian approximation itself we do not have to deal with the  
three-dimensional patologies of the 
contact interaction . So that this scheme 
provides a clear comparison between these 
three different results.
\end{abstract}

\section{Introduction}
\vskip 0.7cm

The non-relativistic interacting Bose gas is 
certainly among the most interesting problems in many-body physics. 
This system has been extensively investigated since fifties motived 
strongly by the superfluid properties of helium \cite{DAPi}. The 
recent successful achiement of Bose-Einstein condensation of alkalis 
vapors \cite{BEC} has renewed a great interest in theoretical studies 
of Bose gases \cite{Dal}.

For dilute gases the macroscopic condensate function is given by the 
Gross-Pitaevskii equation. The effects of noncondensate atoms can be 
calculated in the loop expansion in power of $\sqrt{\rho a^{3}}$, 
where $\rho$ is the density and $a$ is the $S$-wave scattering 
\cite{Hu}. This perturbative method tends to be less reliable in the 
circumstance when higher trap densities are involved. 
On the other hand, non-perturbative 
approximation schemes have been developed in connection with the problem 
of self-interacting, relativistic Bose fields, which became relevant 
for inflationary models \cite{inflation}, and in the analysis of the 
transient phenomena in the collision of the complex nuclear system 
\cite{collision}. This problem has also been used as a testing ground for 
non-perturbative methods which could then be applied to more complicated 
systems of interacting fields. In this context, the Gaussian variational 
approach \cite{variational} has received considerable attention, also in 
view of its relation to extended mean field methods traditionally 
employed in many-body physics \cite{methods1,methods2}.

The main objective of this work is to 
compare the standard perturbation 
results with the unperturbative ones obtained with the 
Gaussian approximation in the case of a 
system of non-relativistic bosons in one 
dimension interacting through 
repulsive, two-body contact interactions. 
The use of an one-dimensional problem gives the 
advantage of dealing with a repulsive contact 
interaction without getting 
into problem due to divergencies \cite{PP,KP}. Also 
the existence of an exact solution in 
this case show us clearly when 
the perturbation theory breaks down and 
how the upperbound non-perturbative 
results stands in this region. 

The Gaussian mean-field approximation which we will employ here 
is based in terms of a time-dependent
projection approach developed earlier for the nonrelativistic nuclear
many-body dynamics by Nemes and de Toledo Piza \cite {TPN}. This scheme 
allows one to formulate a mean-field expansion for the dynamics
of the two-point correlation function from which one recovers the
results of the Gaussian mean-field approximations in lowest order,
i.e., this permits to include and to evaluate higher dynamical
corrections effects to the simplest Gaussian mean-field
approximation. Moreover, the expansion is energy-conserving (for
closed system) to all orders \cite {BFN}. The method was
recently applied in the context of the field-theoretical models, 
namely, to treat an uniform relativistic (1+1)-dimensional 
self-interacting 
boson system described by the $\lambda \phi^{4}$ theory \cite {methods2}, 
to treat an uniform relativistic (1+1)-dimensional self-interacting 
fermion system described by the chiral Gross-Neveu model \cite {Natti1}, 
and  to treat an uniform (3+1)-dimensional relativistic interacting 
boson-fermion system described by the Plasma Scalar model \cite {Erica1}. 

Although we developed a temperature dependent formalism in this 
paper we will concentrate on the zero temperature case and 
calculate in Gaussian mean-field approximation the physical 
quantities related to the optimized free energy such as ground state 
energy and sound velocity to compare with pertubative and exact 
results given in literature \cite{LW,LB}. 
In Section 2, we define the problem and show its exact solution. 
Section 3 reviews the approximation used here and discusses the 
equilibrium solutions for the resulting variational equations. 
Through a truncated version 
of the Gaussian variational equation we obtain the perturbative results 
in Section 4. In Section 5 we show the numerical results obtaneid from the 
Gaussian and perturbative approximations and compare these with the exact 
calculations. Section 6 discusses the range of validity of these 
theoretical methods and main conclusions.  

\section{The problem and the exact solution}

We consider an uniform and isotropic 
system of non-relativistic, interacting, spinless 
bosons in one-dimension described by the Hamiltonian, 

\begin{equation}
H= -\sum_{i} \left(\frac{\partial^{2}}{\partial x_{i}^{2}}\right) +  2c 
\sum_{i,j} \delta(x_{i}-x_{j})
\end{equation}

\noindent with periodic boundary condition in region $0 \leq x_{i} 
\leq L$, 
where $2 c > 0$ is the amplitude of the repulsive delta function 
potential. The only dimensionless intensive variable in the theory is 

\[
\gamma = \frac{c}{\rho}\;\;,
\]

\noindent
where $\rho = N/L$ is the density system. 

Lieb and Liniger in Ref. \cite{LW} have shown that the exact ground state 
energy can be obtained by solving an inhomogeneous Fredholm equation of 
the second kind. So that in the thermodynamic limit ($N,L \rightarrow 
\infty$ such that $\rho =$ fixed constant) we can write the ground state 
energy per particle as

\begin{equation}
\frac{E_{0}}{N} = \frac{1}{\rho} \int_{-K}^{K} f(k) k^{2} dk\;\;\;,
\end{equation}

\noindent where $f(k)$ is the solution of

\begin{equation}
2 c \int_{-K}^{K} \frac{f(p)}{c^2 + (p-k)^{2}} dp = 2 \pi f(k) - 1
\end{equation}

\noindent 
with the condition

\begin{equation}
\int_{-K}^{K} f(k) dk = \rho\;\;\;.
\end{equation}

\noindent 
For computational purposes we can change variables as follows
\[
k = K x \;\; ; \;\; c = K \lambda \;\; ; \;\; f(Kx) = g(x)\;\; ,
\]

\noindent 
in terms of which the Eqs.(2), (3) and (4) become, respectively,

\begin{equation}
\frac{E_{0}}{N} = \rho^{2} \frac{\gamma^{3}}{\lambda^{3}} 
\int_{-1}^{1} g(x) x^{2} dx
\end{equation}

\begin{equation}
1 + 2 \lambda \int_{-1}^{1} \frac{g(x)}{\lambda^{2} + (x-y)^2} 
dx = 2 \pi g(y)
\end{equation}

\begin{equation}
\gamma \int_{-1}^{1} g(x) dx = \lambda\;\;\;.
\end{equation}

\noindent 
For a fixed $\lambda$ we can calculate $g(x)$ having 
$\lambda(\gamma)$ and with it follows $\frac{E_{0}}{N} (\gamma)$ 
and $K(\gamma)$  \cite{LW}. Once we have the ground state energy 
it is possible to calculate the chemical potential 

\begin{equation}
\mu = \frac{\partial E_{0}}{\partial N}\;\;\;. 
\end{equation}
\vskip 0.3cm
\noindent 
From $\mu$ we can calculate the sound velocity derived from the 
macroscopic compressibility

\begin{equation}
v_{s} = 2\left(\mu - \frac{1}{2} \gamma \frac{\partial \mu}
{\partial \gamma} \right)^{1/2}\;\;\;.
\end{equation}
\vskip 0.3cm
\noindent 
Lieb, in another work \cite{LB}, also has shown that this 
theory exhibit 
two types of excitations. The energy of the first type of 
excitation is given by 

\begin{equation}
\epsilon_{1}(p) = -\mu + q^2 + 2 \int_{-K}^{K} k J(k) dk\;\;\;,
\end{equation}
\vskip 0.2cm
\noindent 
where $\mu$ is the chemical potential and  

\begin{equation}
p = q + \int_{-K}^{K} J(k) dk\;\;\;,
\end{equation}
\vskip 0.2cm
\noindent 
while $J(k)$ is again a solution of a Fredholm equation

\begin{equation}
2 \pi J(k) = 2 c \int_{-K}^{K} \frac{J(r) dr}{c^2 + (k -r)^{2}} 
- \pi + 2 \tan^{-1} (q-k)\;\;\;.
\end{equation}
\vskip 0.2cm
\noindent 
The energy of the second type of excitation have the same 
structure and is such that

\begin{equation}
\epsilon_{2}(p) = \mu - q^2 + 2 \int_{-K}^{K} k G(k) dk
\end{equation}

\begin{equation}
p = -q + \int_{-K}^{K} G(k) dk
\end{equation}

\begin{equation}
2 \pi G(k) = 2 c \int_{-K}^{K} \frac{G(r) dr}{c^2 + (k -r)^{2}} 
+ \pi - 2 \tan^{-1} (q-k)\;\;\;.
\end{equation}
\vskip 0.3cm
\noindent 
To solve these equations we use the same procedure used in the 
ground state energy. Here $q$ plays the same role as $\lambda$ in the 
other case.

\section{Gaussian approximation}

In second quantization, rewriting the Hamiltonian (1), in momentum 
representation with periodic boundary conditions in region $L$, follows

\begin{equation}
H=\sum_{\vec{k}} e(k) a_{\vec{k}}^{\dag} a_{\vec{k}}  +
\frac{c}{L} \sum_{\vec{k_{1}} \vec{k_{2}} 
\vec{q}}^{} a_{\vec{k_{1}}+\vec{q}}^{\dag} a_{\vec{k_{2}}-
\vec{q}}^{\dag} a_{\vec{k_{2}}} a_{\vec{k_{1}}}\;\;\;, 
\end{equation}

\noindent 
where $e(k)=k^2$ is the free particle kinetic energy, and the contact 
repulsive ($c > 0$) interaction potential between a pair of 
particles is $2 c \;\delta(\vec{x} - \vec{x'})$. The 
$a_{\vec{k}}^{\dag}$ and $a_{\vec{k}}$ are boson creation 
and annihilation operators, which satisfy the standard boson
commutation relations at equal times. 

\subsection{Projection technique and approximation scheme}

In this section we introduce the time-dependent projection technique
\cite{TPN} which will permit to obtain closed approximations to the
dynamics of a system. It has been developed earlier in the
context of nonrelativistic nuclear many-body dynamics and was recently
applied in the quantum-field theoretical context \cite{methods2,Natti1,
Erica1}. It allows for the formulation of a mean-field expansion for
the dynamics of the two-point correlation function from which one
recovers the results of the Gaussian mean-field approximations in
lowest order. If carried to higher orders it allows for the inclusion
and evaluation of higher dynamical correlation corrections to the
simplest mean-field approximation.
 
We begin by describing the quantum state of the system in Heisenberg 
picture in terms of a many-body density operator $\cal F$, a time
independent, non-negative, Hermitian operator with unit trace. 
$\cal F$ may in general involve statistical 
mixtures, in particular in the
boson number operator, in the spirit of a grand canonical description.
Our implementation of the Gaussian approximation will consist in 
decomposing the full density ${\cal F}$ as 

\begin{equation}
{\cal F}={\cal F}_{0}(t)+{\cal F}'(t)\;\;,
\end{equation}

\vskip 0.5cm

\noindent where ${\cal F}_{0}(t)$ is a Gaussian ansatz which achieves
a Hartree-Fock factorization of traces involving more than two field
operators. The most general hermitian Gaussian density ${\cal F}_{0}(t)$ 
have the form of a exponential of linear and bilinear expressions in the
fields normalized to unit trace \cite{DC,BV}. In the momentum basis, 
it reads

\vskip 0.5cm
\begin{equation}
{\cal F}_{0}=\frac{\exp \left(\sum_{(\vec{k_{1}},\vec{k_{2}})}
\left[A_{\vec{k_{1}},\vec{k_{2}}}a^{\dag}_{\vec{k_{1}}}a_{\vec{k_{2}}}+
\left(B_{\vec{k_{1}},\vec{k_{2}}}a^{\dag}_{\vec{k_{1}}}
a^{\dag}_{\vec{k_{2}}} + \mbox{h.c.} \right) + 
\left(C_{\vec{k_{1}}}a^{\dag}_{\vec{k_{1}}} 
+ \mbox{h.c.} \right)\right]\right)}
{Tr \left\{\exp \left(\sum_{(\vec{k_{1}},\vec{k_{2}})}
\left[A_{\vec{k_{1}},\vec{k_{2}}}a^{\dag}_{\vec{k_{1}}}a_{\vec{k_{2}}}+
\left(B_{\vec{k_{1}},\vec{k_{2}}}a^{\dag}_{\vec{k_{1}}}
a^{\dag}_{\vec{k_{2}}} + \mbox{h.c.} \right) + 
\left(C_{\vec{k_{1}}}a^{\dag}_{\vec{k_{1}}} 
+ \mbox{h.c.} \right)\right]\right)\right\}}\;,
\end{equation}

\vskip 0.5cm

\noindent where the matrix $A_{\vec{k_{1}},\vec{k_{2}}}$ is hermitian. 
The parameters in Eq.(18) are fixed by requiring that mean
values in ${\cal F}_{0}$ of expressions that are bilinear in the
fields reproduce the corresponding ${\cal F}$ averages [see Eqs.(21)
and (22) below]. 

The uniformity and isotropy assumptions we make allow us to restrict 
this general ansatz so that $A_{\vec{k_{1}},\vec{k_{2}}}$ is diagonal, 
$B_{\vec{k_{1}},\vec{k_{2}}}$ vanishes unless $\vec{k_{1}}=-\vec{k_{2}}$ 
and both of these matrices and the $C_{\vec{k_{1}}}$ depend only on the 
magnitudes of the momentum vectors. Finally, by a general canonical 
transformation of the Bogolyubov type, the time-dependent matrix 
${\cal F}_{0}$ can be diagonalized \cite{methods2}. In this case, 
the diagonalization is achieved by defining transformed boson operators 
as

\[
\eta_{\vec{k}} = x_{k}^{\ast} b_{\vec{k}} + y_{k}^{\ast}
b_{-\vec{k}}^{\dag}
\]
\[
\eta_{\vec{k}}^{\dag} = x_{k} b_{\vec{k}}^{\dag} + y_{k}
b_{-\vec{k}}
\]

\noindent 
where

\[
b_{\vec{k}} = a_{\vec{k}} - \Gamma_{k}
\]

\[
\Gamma_{k} =\langle  a_{\vec{k}} \rangle\;\;\;. 
\]

\vskip 0.2cm

\noindent 
We also have used the isotropy of the uniform system to
make the c-number transformation parameters $x_{k}$, $y_{k}$ and
$\Gamma_{k}$ dependent only on the magnitude of $\vec{k}$. In order
for this transformation to be canonical we have still to impose on
the $x_{k}$ and $y_{k}$ the usual normalization condition

\begin{equation}
|x_{k}|^2 - |y_{k}|^2 = 1\;\;.
\end{equation} 
\vskip 0.2cm

In terms of the Bogolyubov quasi-boson operators, the trace-normalized 
Gaussian density operator is now written explicitly as

\begin{equation}
{\cal F}_{0} =  \prod_{\vec{k}} \frac{1}{1 + \nu_{k}}
\left(\frac{\nu_{k}}{1 + \nu_{k}} \right)
^{\eta_{\vec{k}}^{\dag} \eta_{\vec{k}}}\;\;.
\end{equation}

\noindent 
Straightforward calculation shows that

\begin{equation}
Tr(\eta_{\vec{k_{1}}}^{\dag} \eta_{\vec{k_{2}}} {\cal F}_{0})=
\nu_{k_{1}} \delta (\vec{k_{1}} - \vec{k_{2}})
\end{equation}

\vskip 0.2cm 

\noindent 
so that the $\nu_{k}$ are positive quantities corresponding
to mean occupation numbers of the $\eta$-bosons. One also finds that

\begin{equation}
Tr[(x^{*}_{k}a_{\vec{k}}+y^{*}_{k}a_{\vec{-k}}^{\dag})^n {\cal F}_{0}]=
Tr[(\eta_{\vec{k}}+A_{k})^n {\cal F}_{0}]= (A_{k})^n
\end{equation}

\vskip 0.3cm 

\noindent 
with $A_{k}=x_{k}^{*} \Gamma_{k}+y_{k}^{*} \Gamma_{k}^{*}$
so that non vanishing values of the $\Gamma_{k}$ correspond to coherent
condensates of unshifted Bogolyubov transformed bosons. 
We again invoke the uniformity of the system to impose

\begin{equation}
\Gamma_{k}=\delta_{k,0} \Gamma_{0}
\end{equation}
\vskip 0.1cm 

\noindent 
in the calculations to follow.

The ``remainder'' density ${\cal F}^{\prime}(t)$, defined by Eq.(17), 
is a traceless, pure correlation
density. As already remarked, a crucial point to observe is that
${\cal F}_{0}(t)$ can be written as a time-dependent projection of
${\cal F}$, i.e.,

\begin{eqnarray}
{\cal F}_{0}(t) = {\cal P}(t){\cal F}\;\;\;{\rm with}\;\;\;
{\cal P}(t){\cal P}(t)={\cal P}(t)\;.
\end{eqnarray}
\vskip 0.4cm

\noindent In order to completely define this projector we require
further that it satisfies

\begin{equation}
i \dot {\cal F}_{0}(t)=\left[{\cal P}(t),{\cal L}\right]{\cal F}=
\left[{\cal F}_{0}(t),H\right]+{\cal P}(t)\left[H,{\cal F}\right]\;,
\end{equation}

\vskip 0.4cm

\noindent where ${\cal L}$ is the Liouvillian defined as

\begin{equation}
{\cal L}\;\cdot=[H,\;\cdot\;]\;,
\end{equation}

\vskip 0.4cm

\noindent $H$ being the Hamiltonian of the field. Eq.(25) is just the
Heisenberg picture counterpart of the condition $ \dot{{\cal P}}(t)
{\cal F} = 0$ which has been used to define ${\cal P}(t)$ in the
Schr\"odinger picture \cite{BFN}.  It is possible to show that
conditions (24) and (25) make ${\cal P}(t)$ unique and to obtain an
explicit form for this object in terms of the quasi-boson operators
and of the natural orbital occupations see Refs. 
\cite{methods2,TPN,Erica1}. 

The existence of the projector ${\cal P}(t)$ allows one to obtain an
equation relating the correlation part ${\cal F}^{\prime}(t)$ to the
Gaussian part ${\cal F}_0(t)$ of the full density. This can be
immediately obtained from Eqs.(17), (24) and (25) and reads

\begin{equation}
\left(i\partial_{t}+{\cal P}(t){\cal L}\right){\cal F}'(t)=
\left({\cal I}-
{\cal P}(t)\right){\cal L}{\cal F}_{0}(t)\;.
\end{equation}

\vskip 0.5cm

\noindent This equation has the formal solution

\begin{equation}
{\cal F}'(t)={\cal G}(t,0){\cal F}'(0)-i\int_{0}^{t} dt'\;{\cal
G}(t,t') \left({\cal I}-{\cal P}(t')\right){\cal L}{\cal
F}_{0}(t')\;,
\end{equation}

\vskip 0.5cm

\noindent where the first term accounts for initial correlations
possibly contained in ${\cal F}$. The object ${\cal G}(t,t^{\prime})$
is the time-ordered Green's function

\begin{equation}
{\cal G}(t,t')= 
T\left(\exp{\left[i\int_{t'}^{t}d
\tau{\cal P}(\tau){\cal L}\right]}\right)\;.
\end{equation}

\vskip 0.5cm

We see thus that ${\cal F}^{\prime}(t)$, and therefore also ${\cal F}$
[see Eq.(17)], can be formally expressed in terms of ${\cal
F}_{0}(t^{\prime})$ (for $t^{\prime}\leq t$) and of initial
correlations ${\cal F}^{\prime}(0)$. This allows us to express also
the dynamics of the system as functionals of ${\cal
F}_{0}(t^{\prime})$ and of the initial correlations.  Since, on the
other hand, the reduced density ${\cal F}_{0}(t^{\prime})$ is
expressed in terms of the one-boson densities alone, we see that the
resulting equations are essentially closed equations. Note,
however, that the complicated time dependence of the field operators
is explicitly probed through the memory effects present in the
expression (28) for ${\cal F}^{\prime}(t)$. Approximations are
therefore needed for the actual evaluation of this object.  A
systematic expansion scheme for the memory effects has been discussed
in Refs. \cite{methods2,TPN,BFN,Erica1}.  
The lowest order correlation corrections to the
pure mean field approximation, in which ${\cal F}'$ is simply ignored,
correspond to replacing the full Heisenberg time-evolution of
operators occurring in the collision integrals by a mean-field
evolution governed by 

\[
H_{0}={\cal P}^{\dag}(t)H\;\;.\nonumber
\]

\vskip 0.3cm

\noindent Consistently with this approximation, ${\cal L}$ is replaced
in (25) and (26) by ${\cal L}_{0}\;\cdot=[H_{0},\;\cdot\;]$.  In this
way correlation effects are treated to second order in $H$ in the
resulting collision integrals.

An important feature of this scheme (which holds also for higher
orders of the expansion \cite{BFN}) is that the mean energy is
conserved, namely

\[
\frac{\partial}{\partial t}\langle H\rangle=0
\]

\noindent where

\[
\langle H\rangle=Tr\;H{\cal F}_{0}(t)+Tr\;H{\cal F}^{\prime}(t)\;.
\]

In the following sections we apply the general expressions obtained in
above to treat a uniform boson system described by the Hamiltonian 
given in Eq.(16). We will consider only the lowest
(mean-field) approximation, corresponding to ${\cal F}^{\prime}(t)=0$.
Collisional correlations will be treated elsewhere.

\subsection{Temperature time-dependent Gaussian treatment}

First, it is important to note that the 
truncated density ${\cal F}_{0}$ in
general breaks the global gauge symmetry of $H$, which is responsible for
the conservation of the number of $a$-bosons. It is possible to verify 
that assumption calculating the dispersion

\[
\langle N^{2} \rangle - \langle N \rangle^{2} = Tr [ {\cal F}_{0} 
\sum_{\vec{k},\vec{k'}} a_{\vec{k}}^{\dag} a_{\vec{k}} 
a_{\vec{k'}}^{\dag} a_{\vec{k'}}] - \{Tr[ {\cal F}_{0} \sum_{\vec{k}} 
a_{\vec{k'}}^{\dag} a_{\vec{k'}}]\}^{2}\;\;\;. 
\]

\vspace{0.2cm}     

\noindent  
Calculating the traces above we have

\begin{eqnarray}
\langle N^{2} \rangle - \langle N \rangle^{2} &=&   
2| \Gamma_{0} |^{2} [| x_{0} |^{2} \nu_{0} + |y_{0}|^{2} (1 + \nu_{0})] 
- 2\Gamma_{0}^{\ast^{2}} x_{0} y_{0}^{\ast} (1 + 2 \nu_{0}) -   
\nonumber  \\
&&\nonumber \\
&&-2\Gamma_{0}^{2} y_{0} x_{0}^{\ast} (1 + 2 \nu_{0}) 
+ |\Gamma_{0}|^{2} +\sum_{\vec{k}}[|x_{k}|^{2} \nu_{k} 
+ (1 + \nu_{k}) |y_{k}|^{2} ] +    \nonumber \\
&&+\sum_{\vec{k}} \{ [|x_{k}|^{2} \nu_{k} + (1 + \nu_{k}) |y_{k}|^2]^{2} 
+ |x_{k}|^{2} |y_{k}|^{2} (1+ 2 \nu_{k})^{2} \}\;\;\;.   \nonumber \\
\nonumber
\end{eqnarray}

\noindent 
Furthermore, mean values
of many-boson operators taken with respect to ${\cal F}_{0}$ will
contain no irreducible many-body parts, so that the replacement of
$\cal F$ by ${\cal F}_{0}$ amounts to a mean field approximation. The
states described by ${\cal F}_{0}$ have therefore to be interpreted
as ``intrinsic'' mean field states.

In order to develop a finite temperature treatment
within this framework, we look for variational extrema of an
approximation $\Omega$ to the grand potential written in terms of
the Gaussian density operator ${\cal F}_{0}$ as

\begin{equation}
\Omega=Tr[{\cal F}_{0}(H - \mu N + KT \ln {\cal F}_{0})]
\end{equation}

\noindent 
with

\[
N = \sum_{\vec{k}} a_{\vec{k}}^{\dag} a_{\vec{k}}\;\;\;,
\]

\noindent 
where $K$ is the Boltzmann constant and the Lagrange
multipliers $\mu$ and $T$ will play the role of chemical potential and
temperature respectively. The replacement of the full density operator
$\cal F$ by ${\cal F}_{0}$ implies not only truncation of correlation
energies but also replacement of the full entropy by the mean field
entropy
$S_{0}=-K Tr[{\cal F}_{0} \ln {\cal F}_{0}]$. Variations will be taken
with respect to the parameters which determine the Gaussian density
${\cal F}_{0}$, and the value of $\mu$ is fixed by choosing the
particle density of the system. A detailed justification of this
general procedure has been given \cite{PP,BV}, 
who have shown that it corresponds to the optimal
determination (in the variational sense) of the grand partition
function (hence of the grand potential) for the problem when one
restricts oneself to trial density operators of the form given in
Eq.(20). We have to emphasize that the use of this functional $\Omega$ 
does not assure that we will have the best variational results for other 
observables. See discussion in Secs. 5 and 6.

An important simplification occurs in the case of 
stationary problems. In this case $x_{k}$, $y_{k}$
and $\Gamma_{0}$ can be taken to be real and we can use
a simple parametric representation that automatically satisfies
the canonicity  condition. It reads

\begin{equation}
x_{k} = \cosh \sigma_{k}\;\;\;,\;\;\;y_{k} = \sinh \sigma_{k}\;\;. 
\end{equation} 

\vskip 0.3cm

\noindent 
It is then straightforward to evaluate the traces involved
in $\Omega$ to obtain

\begin{eqnarray}
\Omega &=&\sum_{\vec{k}} {\left(e(k) - \mu + \frac{4c \Gamma_{0}^{2}}
{L}\right)\left[\frac{(1 + 2 \nu_{k})
\cosh 2\sigma_{k} -1}{2} \right]}-\mu \Gamma_{0}^{2}+
\frac{c \Gamma_{0}^{4}}{L}  \nonumber \\
&&-\frac{c \Gamma_{0}^{2}}{L} \sum_{\vec{k}}^{}
{(1 + 2 \nu_{k})\sinh 2 \sigma_{k} }+
\frac{2c}{L}\left \{\sum_{\vec{k}}\left[ \frac{(1+
2 \nu_{k})\cosh 2\sigma_{k} -1 }{2}\right] \right\}^{2}  \nonumber \\
&&+\frac{c}{4 L} \left\{\sum_{\vec{k}} {(1 + 2 
\nu_{k}) \sinh 2 \sigma_{k} } \right\}^{2}  \\
&&-KT \sum_{\vec{k}} \left[(1+\nu_{k})\ln(1+\nu_{k})-
\nu_{k}\ln\nu_{k} \right]\;\;\;. \nonumber  
\end{eqnarray}

\noindent  
In a similar way the number constraint 
$Tr[N{\cal F}_{0}]=\langle N \rangle$ evaluates to

\begin{equation}
\langle N \rangle = \Gamma_{0}^{2} + \sum_{\vec{k}}^{}
\left[\frac{(1 + 2 \nu_{k}) \cosh2 \sigma_{k}  -1}{2} \right]\;\;\;. 
\end{equation}

\subsection{Equilibrium solutions}

Equations determining the form of the Gaussian density
${\cal F}_{0}$ appropriate for thermal equilibrium are in
general derived by requiring  $\Omega$, given by (32), to be 
stationary under arbitrary variations of $\Gamma_{0}$,
$\sigma_{k}$ and $\nu_{k}$. Variation with respect to
$\Gamma_{0}$ gives the gap equation

\begin {equation}
\Gamma_{0} \left\{\frac{4 c}{L} \Gamma_{0}^{2} - 2 \mu - 
4 c A + 8 c B \right\} =0\;\;\;,
\end{equation} 

\vskip 0.4cm

\noindent 
where we define in the thermodynamic limit, 
$\sum_{k} \rightarrow \frac{L}{2 \pi} \int_{-\infty}^{+\infty} dk $,
the quantities $A$, $B$ and $C$ given by

\begin{eqnarray}
A &=& \frac{1}{4 \pi} \int_{-\infty}^{\infty} (1 + 2 \nu_{k}) 
\sinh 2\sigma_{k} \: dk  
\nonumber \\
\nonumber \\
B &=& \frac{1}{4 \pi} \int_{-\infty}^{\infty} [(1 + 2 \nu_{k}) 
\cosh 2\sigma_{k} -1] \: dk 
\nonumber \\
\nonumber \\
C &=& \frac{1}{4 \pi} \int_{-\infty}^{\infty} e(k)
[(1 + 2 \nu_{k}) \cosh 2\sigma_{k} -1] \: dk\;\;\;. 
\nonumber
\end{eqnarray}

\vskip 0.3cm

The gap equation, Eq.(34), admits a
solution with a non-vanishing value of $\Gamma_{0}$ (condensed phase)
obtained by requiring that the expression in curly brackets vanishes.
This condensed phase solution 
involves the number constraint, Eq.(33), in addition
to the values of $\nu_{k}$ and $\sigma_{k}$, which are determined
by the remaining variational conditions on $\Omega$. In order to
simplify the algebraic work involved in the study of this class of
solutions, it is sometimes convenient to use the Eq.(33)  
to eliminate $\Gamma_{0}$ from $\Omega$, which then becomes

\[
\Omega = F - \mu \langle N \rangle\;\;.
\]

\noindent 
This identifies the free energy $F$ as 

\begin{eqnarray}
F&=&LC-2c \rho (A-B)L+c(\rho^{2}+A^{2}-B^{2})L
+2cABL \nonumber \\\\
& &- \frac{LKT}{2 \pi} \int_{-\infty}^{\infty} 
\left[ (1+\nu_{k})\ln(1+\nu_{k})-\nu_{k}
\ln\nu_{k} \right] \: dk \;\;\;.\nonumber 
\end{eqnarray}
\vskip 0.2cm
\noindent 
For condensed  phase, the chemical potential follows from Eqs.(33) 
and (34)

\begin{equation}
\mu = 2 c (\rho - A + B)\;\;.
\end{equation}
\vskip 0.2cm
\noindent 
Extremizing $F$ by setting derivatives with respect to $\sigma_{k}$ 
and $\nu_{k} $ equal to zero one gets

\begin{equation}
\tanh 2 \sigma_{\vec{k}} = \frac{2 c [\rho -B-A]}{e(k) +
2 c [\rho -B+A]} 
\end{equation}

\noindent 
and

\begin{equation}
\nu_{k} =\frac{1}{\{exp[\sqrt{\Delta}/{KT}]-1\}} \;\;,
\end{equation}

\noindent 
where

\begin{equation}
e_{g}(k) = \sqrt{\Delta} = \sqrt{e(k)^{2} + 4 c e(k) [\rho - B + A] +
16 c^2 [\rho -B] A}
\end{equation}
\vskip 0.3cm
\noindent 
corresponds to the excitation spectrum in our approximation. 
This result can be confirmed through an RPA calculation Ref. \cite{PT}. 
Finally, $A$ and $B$ can be found from gap equations (37-38), for given 
values of $\rho$ and $T$, by solving numerically the system of 
equations

\begin{eqnarray}
A &=&\frac{1}{4 \pi} \int_{-\infty}^{\infty}
\left\{ \frac{2 c [\rho -B-A]}{\sqrt{\Delta}}\left[
1+\frac{2}{\{exp[\sqrt{\Delta}/{KT}]-1\}}\right] \right\}dk 
\nonumber \\\\
B &=&\frac{1}{4 \pi} \int_{-\infty}^{\infty}
\left\{ \frac{e(k)+2 c [\rho -B+A]}{\sqrt{\Delta}}
\left[1+\frac{2}{\{exp[\sqrt{\Delta}/{KT}]-1\}}\right]
-1 \right\}dk\;\;\;. \nonumber
\end{eqnarray} 
\vskip 0.3cm

\noindent
Therefore, in terms of the values of $A$ and $B$, we can calculate 
$C$ and the thermodynamic functions also numerically.

On the other hand, the gap equation (34) may also admit, 
besides the condensed phase, 
the trivial solution $\Gamma_{0} = 0$, which corresponds to a 
non-condensed phase. For this phase, we extremize $\Omega$, given  
in Eq.(32), by setting derivatives with respect to $\sigma_{k}$ 
and $\nu_{k} $ equal to zero one gets

\begin{equation}
\tanh 2 \sigma_{k} = \frac{-2 c A}{e(k) - \mu + 2 c B}
\end{equation}
\vskip 0.3cm

\noindent and

\begin{equation}
\nu_{k} =\frac{1}{\{exp[\sqrt{\Delta}/{KT}]-1\}} \;\;,
\end{equation}

\noindent 
where

\begin{equation}
e_{g}(k) = \sqrt{\Delta} = e(k) - \mu + 4 c \rho \;\;.
\end{equation}
\vskip 0.3cm

\noindent
For non-condensed phase ($\Gamma_{0} = 0$) 
the only possible solution, from gap equations (41-42),
is $A=0$ following that $\sigma_{k}=0$. 
Therefore, from Eqs.(33) and (42-43), we have the solution

\begin{equation}
\rho = \frac{1}{2 \pi} \int_{-\infty}^{\infty} \frac{dk}{\exp[(k^2 
- \mu + 4 c \rho)/KT]-1}\;\;\;,
\end{equation}

\vskip 0.3cm

\noindent 
which serves to determine $\mu$. That is the standard Hartree-Fock 
result which corresponds to a shift in 
the chemical potential when compared with the ideal Bose gas \cite{RK} . 
In this phase, we show in Ref. \cite{PT} that when 
$T \rightarrow 0$ we have $\mu \rightarrow 4 c \rho$.

Particularly in this paper we are interested on properties 
for $T=0$, since in this case we can compare our results with 
perturbative and exact solutions existent in the 
literature \cite{LW,LB}. When $T=0$, the ground state energy  
is then given by

\begin{equation}
\frac{F(T=0)}{N} = \frac{E}{N} = \frac{C}{\rho} - 2 c (A-B) 
+ \frac{c}{\rho}(\rho^{2} + A^{2} - B^{2}) + 2 \frac{c}{\rho} A B\;\;\;.
\end{equation}
\vskip 0.3cm

We still have to decide what phase shall we use when $T=0$. To do so,  
let us examine the chemical potencial $\mu$ since from Eq.(36). When 
$A>0$ follows from Eq.(36) that 
$\mu_{\Gamma_{0}\ne 0} < 4 c \rho = \mu_{\Gamma_{0}=0}$ and 
it is easy to check that for a given $\rho$ at $T=0$ there are always $A$ 
and $B$ that are solutions of the system (40). This does not happen in 
the case of non-condensed phase at $T=0$ \cite{PP}. Therefore, we 
conclude that the stable phase at $T=0$ for any 
$\rho$ is the one with $\Gamma_{0} \neq 0$.

Finally, in this calculation $A$,$B$ and $C$ are finite meaning that no 
renormalization is needed. This is a quite different result from the 
three-dimensional case Ref. \cite{PP,PT}

\section{Independent $\eta$-bosons - Perturbative results}

In Ref. \cite{PP} it is shown that from a convenient truncation of the
Gaussian approximation in the three-dimension problem, it is possible to 
obtain the usual perturbative results in the parameter $2c$. 
Here we will follow the discussion of Ref.\cite{LW,LB}.

This truncation of the Gaussian approximation consists in neglecting all 
terms representing interactions between $\eta$-bosons, what amounts to 
dropping all double integrals in Eq.(35). The variational conditions on 
$\sigma_{k}$ and $\nu_{k}$ 
appear then as

\begin{equation}
\tanh 2 \sigma_{k} = \frac{2c \rho}{k^{2} + 2c \rho}  
\end{equation}

\noindent 
and

\begin{equation}
\nu_{k} =\frac{1}{\{exp[\sqrt{k^{4}+4 c \rho k^{2}}/{KT}]-1\}} \;\;,
\end{equation}
\vskip 0.3cm

\noindent 
which gives us the usual phonon excitation spectrum

\begin{equation}
\epsilon(p) = \sqrt{p^4 + 4 c \rho p^2}\;\;\;.
\end{equation}
\vskip 0.2cm

\noindent 
When we using these results for calculating $F$, again stressing 
that there is no need for regularization here, we obtain

\begin{equation}
\frac{E}{N} = \rho^{2} \gamma \left(1 
- \frac{4 \sqrt{\gamma}}{3 \pi} \right)\;\;.
\end{equation}
\vskip 0.3cm

\noindent 
Using Eq.(8) and Eq.(9) for the chemical potential and the sound 
velocity we get

\begin{equation}
\mu = 2 \gamma \left(1 - \frac{\sqrt{\gamma}}{\pi} \right) {\rho}^{2}
\end{equation}

\begin{equation}
v_{s} = 2 \rho \left[\left(\gamma - \frac{1}{2 \pi} \gamma^{\frac{3}{2}} 
\right)\right]^{\frac{1}{2}}\;\;\;.
\end{equation}
\vskip 0.5cm
                                                                          
\noindent 
We can check that the sound velocity 
obtained from the excitation spectrum Eq.(48) defined by

\begin{equation}
v_{s} = lim_{p \rightarrow 0} \frac{\partial \epsilon(p)}{\partial p}
\end{equation}
\vskip 0.3cm
\noindent gives us   $v_{s} = 2 \rho \sqrt{\gamma}$ , that is exactly one 
order lower of the one obtained through the compressibility Eq.(51).

\section{Numerical Results}

We did all our calculations for a fixed $\rho = 1$. 
For the exact solution
we follow the steps of Ref.\cite{LW} 
solving numerically the Eqs.(5),(6) 
and (7). For the Gaussian approximation we 
solve the system of Eqs.(39) and (40) taking $T=0$, 
and we calculate the ground state energy given by the Eq.(45). 
Finally for the pertubative 
results we use closed expression (49). 
All this results are shown in Fig.1, 
where we plot the ground state energy as 
a function of $\gamma$. We clearly 
see the upperbound Gaussian results 
and, as expected, satisfactory results for 
low $\gamma$ of both perturbative and Gaussian calculations.

For the exact and Gaussian sound velocity 
we differentiate numerically the 
respective ground state curve. For the 
perturbative solution 
we have two possibilities one using the 
compressibility through Eq.(51) or 
in lower order using the excitation 
spectrum Eq.(48). We plot the sound 
velocity as a function of $\gamma$ in 
Fig.2 . We note that the Gaussian 
result is also an upperbound and for 
this range of $\gamma$ the results 
obtained from perturbative theory 
using the compressibility and the exact 
solution are graphically indistinguishable.

Finally for the excitation spectrum 
we solve numerically (10), (11) and (12) 
for the first type of excitations and 
(13), (14) and (15) for the second 
type. To calculate the excitation 
spectrum in the Gaussian approximation 
using Eq.(39), once we have solved the system (40). For 
the perturbative theory we use Eq.(48). In Fig.3 and Fig.4  we 
plot the excitation energy as a function of the momentum for 
$\gamma = 0.787094$ and $\gamma = 3.07725$ 
respectively. Neither perturbative 
nor Gaussian approximation describe the 
exact second type of excitations and 
the Gaussian approximation introduces a 
gap in the excitation energy that gets 
bigger if we go to the non-perturbative limit.

\section{Conclusion}

In summary, we have used a one-dimensional problem to 
understand , in a clear way, the range of validity of 
the perturbative and the Gaussian variational approximation 
in the context of interacting Bose gas. The methods were  
applied to calculated the grand-potential functional and 
compared to the exact results. In this way 
we avoid problems related to renormalization 
\cite{PP} and have the exact 
solution for the ground state energy and 
excitation spectrum \cite{LW,LB}.

We were able to see in Fig. 1 that the 
ground state energy for low values 
of $\gamma$ both perturbative and Gaussian 
(non-perturbative) results are 
quite good. As $\gamma$ increases we have 
a region where the exact solution 
is in between the two approximations. For 
higher values of $\gamma$ the 
perturbatives results breaks down completelly 
and the Gaussian approximation 
still gives us an good upperbound result. 
This shows us that for high values 
of $\gamma$ a self-consistent non-perturbative 
calculation should be used. 
For the sound velocity, Fig. 2, we see 
that both perturbative and Gaussian 
results are upperbound approximation to the exact result.
From Figs. 3 and 4, we see a very interesting feature of the Gaussian 
approximation, namely, it  produces a coherent macroscopic occupation 
for $k=0$. This occupation is artificial because as we can see from 
the exact solution there is no macroscopic occupation for $k=0$.

Therefore, the results that are related 
directly from the thermodynamic 
potential, such as the ground state energy and 
the sound velocity, obtained through 
the compressibility, are good upperbounds. 
Yet the gap in the excitation 
spectrum, that came from the occupations, 
are very different from the exact 
solution. So, the fact that using the 
optimal determination of the grand 
potential does not assure that we will 
have the best variational result for 
the occupations \cite{BV}. 
Thus, for better results in other 
observables we should involve ourselves 
in a more difficult task of finding 
the best functional related to this specific 
observable with the same trial  
density operator of the form given in Eq.(20).  
Finally we want to point out that the formalism 
develloped here alows us 
to go in straightfoward way to $T \neq 0$ still 
keeping the variational 
caracteristic of having an upperbound result \cite{PP}.

\vskip 0.7cm

\centerline{\Large\bf Acknowledgments}

\vskip 0.7cm

One of the authors (P.R.I.T.) thanks the Conselho Nacional de
Desenvolvimento Cient{\'{\i}}fico e Tecnol\'ogico (CNPq), Brazil.
One of the authors (P.L.N.) was supported during this work by 
Conselho Nacional de Desenvolvimento Cient{\'{\i}}fico e 
Tecnol\'ogico (CNPq), Brazil; and by Funda{\c{c}\~ao} de Amparo 
\`a Pesquisa do Estado de S\~ao Paulo (FAPESP), Brazil.

\section{Figure captions}

Fig.1. The ground-state energy as a  function of $\gamma=c/\rho$. 
The full curve 1 gives the exact solution obtained numerically. 
Curve 2 show the Gaussian approximation also obtained numerically. 
Curve 3 is the result of perturbation theory.

\vskip 0.5cm
\noindent 
Fig.2. The sound velocity $v_{s}$, as a function of $\gamma = c/\rho$. 
Curve 1 is the result obtained from perturbative theory using the 
excitation spectrum. Curve 2 derived from the macroscopic 
compressibility using the Gaussian approximation. Curves 3 and 4, 
that is graphically indistinguishable in this region, are the exact 
result and the perturbation solution obtained 
trought the macroscopic compressibility. 

\vskip 0.5cm
\noindent
Fig.3 The excitation energy as a function of the momentum for 
$\gamma = 0.787094$. The full curves 1 and 2 corresponds to the two 
types of excitations (exact result). Curve 3 corresponds to the 
Gaussian spectrum and curve 4 is the pertubative spectrum.

\vskip 0.5cm
\noindent
Fig.4 The excitation energy as a function of the momentum for 
$\gamma = 3.07725$. The full curves 1 and 2 corresponds to the two 
types of excitations (exact result). Curve 3 corresponds to the  
Gaussian spectrum and curve 4 is the pertubative spectrum.

\end{document}